\documentclass{appolb}
\usepackage{graphicx}

\usepackage{multirow}
\usepackage{ctable}
\usepackage{booktabs}
\usepackage{array}
\usepackage{tabularx}
\usepackage{xcolor}
\usepackage{pstricks}
\usepackage{xspace}

\newcommand{\GeV}{\ensuremath{\mathrm{GeV}}\xspace}

\newcommand{\TeV}{\ensuremath{\mathrm{TeV}}\xspace}

\begin{document}
\title{Production of heavy particle pairs via photon-photon processes at the LHC%
\thanks{Presented at XXVI Cracow EPIPHANY Conference, LHC Physics: Standard Model and Beyond, 7-10 January 2020}%
}
\author{Marta {\L}uszczak
\address{College of Natural Sciences, Institute of Physics, University of Rzesz\'ow,\\ ul. Pigonia 1, PL-35310 Rzesz\'ow, Poland}
\\
{Antoni Szczurek
\address{Institute of Nuclear Physics Polish Academy of Sciences,\\
ul. Radzikowskiego 152, PL-31342 Krak\'ow, Poland}
}
}

\maketitle
\begin{abstract}
We discuss production of $W^+ W^-$ pairs and $t \bar t$ quark-antiquark pairs
in proton-proton collisions induced by two-photon fusion including transverse momenta of incoming photons.
The unintegrated inelastic fluxes (related to proton dissociation) of photons
are calculated based on modern parametrizations of deep inelastic
structure functions in a broad range of $x$ and $Q^2$.
We focus on processes with single and double proton dissociation.
Highly excited remnant systems hadronise producing particles that
can be vetoed in the calorimeter. We calculate associated gap
survival factors. The gap survival factors depend on the process, mass
of the remnant system and collision energy.
The rapidity gap survival factor due to remnant fragmentation for double
dissociative (DD) collisions is smaller than that for single
dissociative (SD) process. We observe approximate factorisation: $S_{R,DD} \approx S_{R,SD}^2$
when imposing rapidity veto. For the $W^+W^-$ final state, the remnant
fragmentation leads to a taming of the cross section when the rapidity
gap requirement is imposed. Also for $t \bar t$ quark-antiquark pairs
such a condition reverses the hierarchy observed for the case when
such condition is taken into account. Our results imply that for the production of such heavy objects
as $t$ quark and $\bar t$ antiquark the virtuality of the photons
attached to the dissociative system are very large ($Q^2 <$ 10$^{4}$
GeV$^2$). A similar effect is observed for the $W^+ W^-$ system.

\end{abstract}
  
\section{Introduction}

Photon-induced processes in proton-proton
interactions have become very topical recently.
Experimentally they can be separated from other competing processes
by imposing rapidity gaps around the electroweak vertex.
Both charged lepton pairs $l^+ l^-$
\cite{Chatrchyan:2011ci,Chatrchyan:2012tv,Aad:2015bwa,Cms:2018het,Aaboud:2017oiq}
and electroweak gauge bosons $W^+ W^-$ \cite{Khachatryan:2016mud,Aaboud:2016dkv}
were recently studied experimentally at the Large Hadron Collider.
In particular, processes with $W^+ W^-$ are of special interest in the context of searches beyond Standard Model \cite{Chapon:2009hh,Pierzchala:2008xc}.
There are, in general, different categories of such processes depending
on whether the proton stays intact or undergoes an electromagnetic
dissociation (see e.g. \cite{daSilveira:2014jla,Luszczak:2015aoa}).

The $W^+ W^-$ production in proton-proton processes via the $\gamma \gamma \to W^+ W^-$ subprocess was studied in collinear \cite{Luszczak:2014mta} and transverse momentum dependent factorisation \cite{Luszczak:2018ntp} approaches.
In our paper \cite{Luszczak:2018ntp} we showed that
rather large photon virtualities and large mass proton excitation are characteristic for the $\gamma \gamma \to W^+ W^-$ induced processes.
Our main aim was to estimate gap survival factor
associated with the remnant hadronisation, which destroys
the rapidity gap. In \cite{Forthomme:2018sxa} we concentrated on the
effect related to remnant fragmentation and its destroying of 
the rapidity gap. 
Finally in \cite{Luszczak:2018dfi} we calculated cross section for the 
photon-photon contribution for the $pp \to t \bar t$ reaction including 
also effects of gap survival probability.

\section{A sketch of the formalism}

\begin{figure}
  \centering
  \includegraphics[width=.32\textwidth]{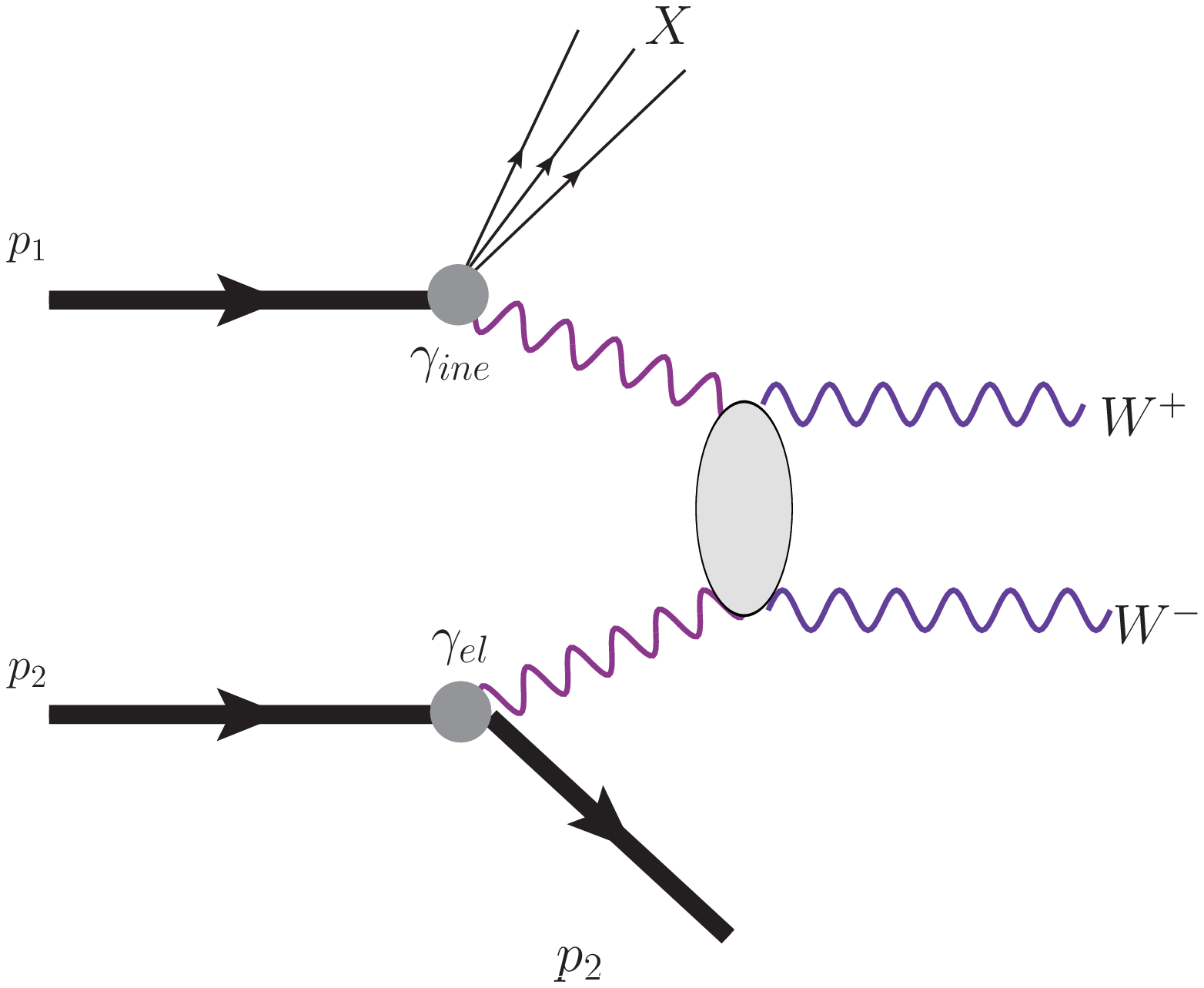}
  \includegraphics[width=.32\textwidth]{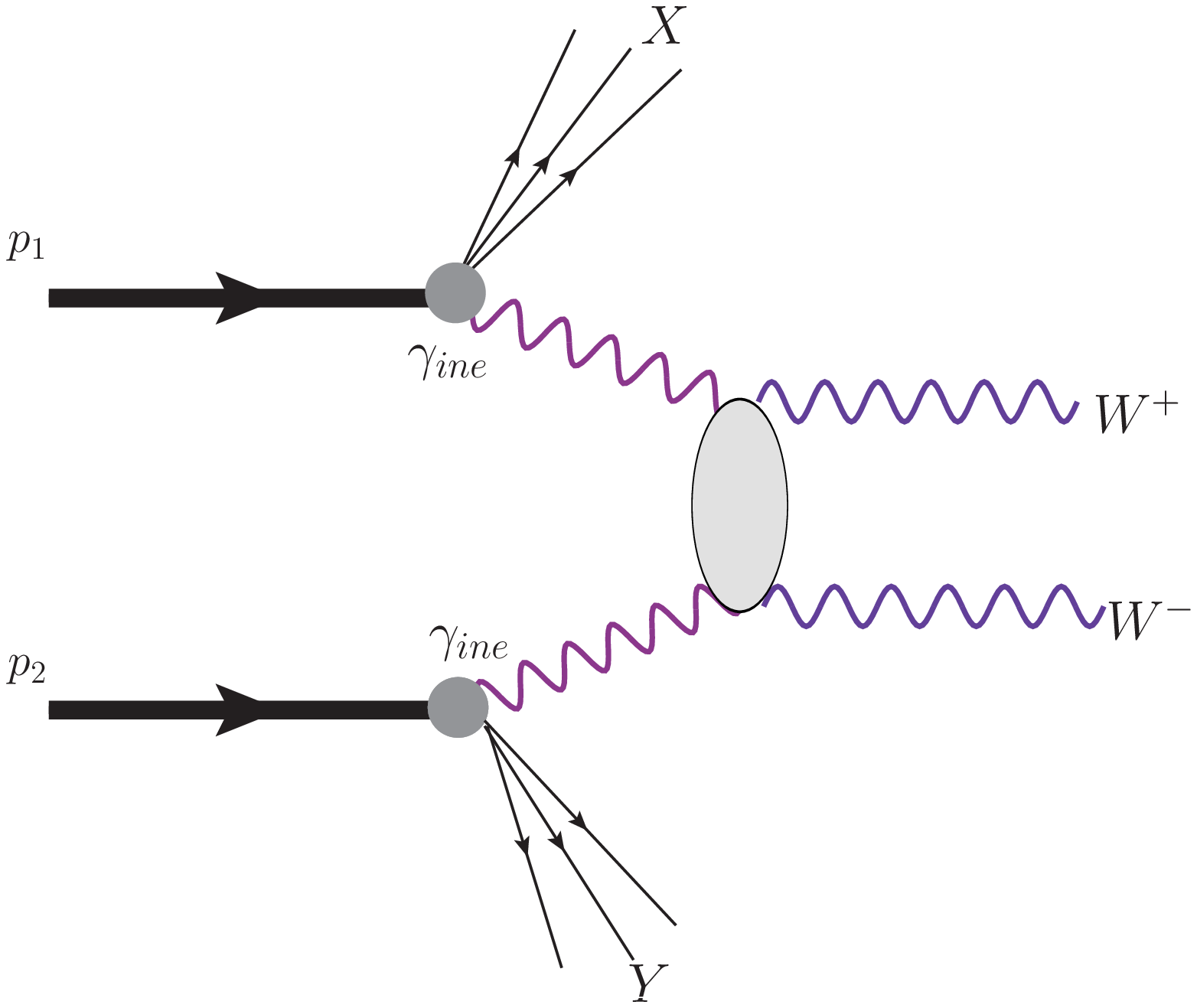}  \\
  \includegraphics[width=.30\textwidth]{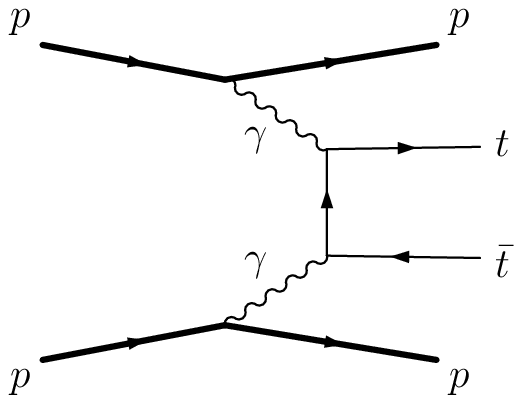}
  \includegraphics[width=.30\textwidth]{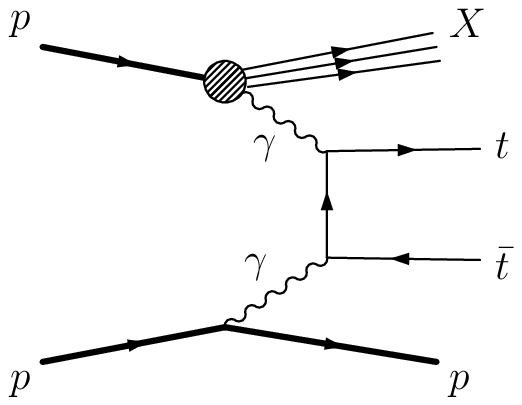}
  \includegraphics[width=.30\textwidth]{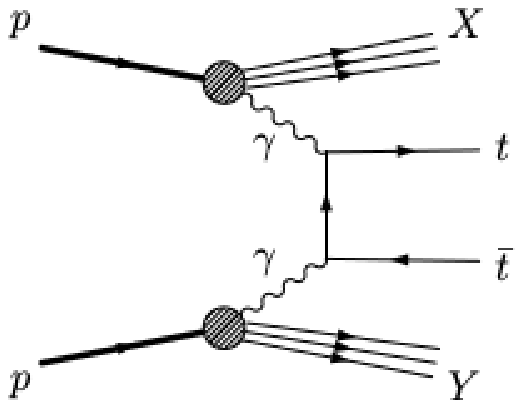}  
  \caption{Diagrams representing different categories of photon-photon induced mechanisms for production of $W^+W^-$ pairs (top panel) and for production of $t \bar t$ pairs (bottom panel).}
\label{fig:diagrams}
\end{figure}

In our analyses of heavy particle pair production via photon-photon 
processes we included different categories of processes shown in
Fig.\ref{fig:diagrams}.
In contrast to other authors our calculations are based on unintegrated 
inelastic photon fluxes.
The unintegrated photon fluxes can be obtained using the following equation:

\begin{eqnarray}
\gamma^p_{\rm in}(x,\vec{q}_T) = {1\over x} 
{1 \over \pi \vec{q}_T^2} \, \int_{M^2_{\rm thr}} dM_X^2 {\cal{F}}^{\mathrm{in}}_{\gamma^* \leftarrow p} (x,\vec{q}_T,M^2_X) \, ,
\label{eq:inelastic}
\end{eqnarray}
and we use the functions $ {\cal{F}}^{\mathrm{in}}_{\gamma^* \leftarrow p}$:
\begin{eqnarray}
{\cal{F}}^{\mathrm{in}}_{\gamma^* \leftarrow p} (x,\vec{q}_T) &=& {\alpha_{\rm em} \over \pi} 
\Big\{(1-x) \Big( {\vec{q}_T^2 \over \vec{q}_T^2 + x (M_X^2 - m_p^2) + x^2 m_p^2  }\Big)^2  
{F_2(x_{\rm Bj},Q^2) \over Q^2 + M_X^2 - m_p^2}  \nonumber \\
&+& {x^2 \over 4 x^2_{\rm Bj}}  
{\vec{q}_T^2 \over \vec{q}_T^2 + x (M_X^2 - m_p^2) + x^2 m_p^2  }
{2 x_{\rm Bj} F_1(x_{\rm Bj},Q^2) \over Q^2 + M_X^2 - m_p^2} \Big\} \, .
\label{eq:flux_in}
\end{eqnarray}
The virtuality $Q^2$ of the photon depends on the photon transverse momentum ($\vec{q}_T^2$) and the proton remnant mass ($M_X$):
\begin{eqnarray}
Q^2 =  {\vec{q}_T^2 + x (M_X^2 - m_p^2) + x^2 m_p^2 \over (1-x)} \, .
\label{eq:qt}
\end{eqnarray}
The proton structure functions $F_1(x_{\rm Bj},Q^2)$ and $F_2(x_{\rm Bj},Q^2)$ depend on:
\begin{eqnarray}
x_{\rm Bj} = {Q^2 \over Q^2+M^2_X -m_p^2}.
\end{eqnarray}
In Eq.~(\ref{eq:flux_in}) we use both $F_2(x_{\rm Bj},Q^2)$ and 
$F_L(x_{\rm Bj},Q^2)$, where
\begin{eqnarray}
F_L(x_{\rm Bj},Q^2) = \Big( 1 + {4 x_{\rm Bj}^2 m_p^2 \over Q^2} \Big) F_2(x_{\rm Bj},Q^2) - 2 x_{\rm Bj} F_1(x_{\rm Bj},Q^2)
\end{eqnarray}
is the longitudinal structure function of the proton.


The photon fluxes enter the 
$p p \to X + (\gamma^\ast \gamma^ast \rightarrow W^+ W^- Y$ and the
$p+p \rightarrow X + (\gamma^\ast \gamma^\ast \rightarrow t \bar t) + Y$
production cross section. Details of the cross sections calculations are
presented in our original papers \cite{Luszczak:2018ntp}, \cite{Forthomme:2018sxa}, \cite{Luszczak:2018dfi}.

\section{Results for $W^+W^-$ pairs production}

Before studying the hadron level we calculated the gap survival
factor on the parton level. In such a case it is the outgoing parton 
(jet or mini-jet), which is struck by the virtual photon and destroys the rapidity gap.

\begin{figure}
\centering
\includegraphics[width=.63\textwidth]{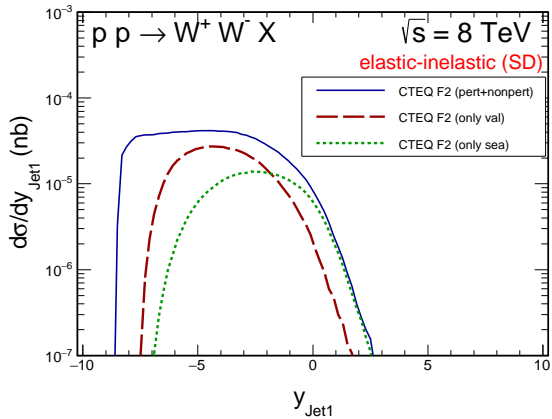}
\caption{Jet rapidity distribution using a LO
partonic distribution at large $Q^2$. The solid line is a sum of all contributions.
The dashed line is for the valence component and the dotted line is for
the sea component.
}
 \label{fig:dsigma_dyjet}
\end{figure}

The gap survival factor can be then defined as:
\begin{equation}
S_R(\eta_{\rm cut}) = 1 - {1 \over \sigma}
\int_{-\eta_{\rm cut}}^{\eta_{\rm cut}} \frac{{\rm d}\sigma}{{\rm d}
    \eta_{\rm jet}} {\rm d} \eta_{\rm jet}, \;
\label{parton_model_gap_survival}
\end{equation}
where ${\rm d}\sigma / {\rm d}\eta_{\rm jet}$ is the rapidity distribution of the cross
section for $W^+ W^-$ production as a function of rapidity of the extra
jet ({\emph de facto} parton) and $\sigma$ is the associated integrated cross
section.
In Fig. \ref{fig:dsigma_dyjet} we show ${\rm d}\sigma /{\rm d}\eta_{\rm jet}$ as a function
of $\eta_{\rm jet}$. No extra cuts are imposed here.
We get a very broad distribution in $\eta_{\rm jet}$ (see solid line).

\begin{figure}
\centering
\includegraphics[width=.63\textwidth]{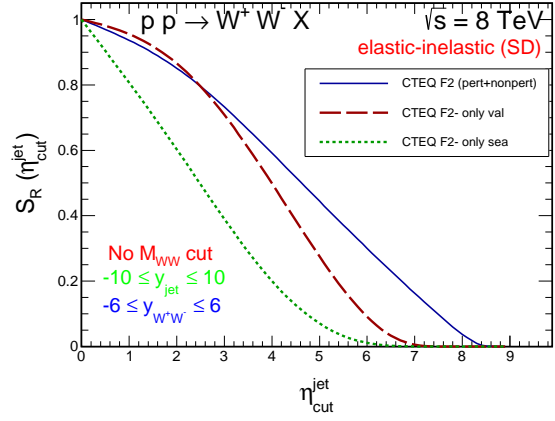}
\caption{Gap survival factor associated with the jet emission. The solid line is for
the full model, the dashed line for the valence contribution
and the dotted line for the sea contribution.
}
 \label{fig:S_R_partonic}
\end{figure}

We also presented the parton level gap survival factor as a function of
the the size of the window $(-\eta_{cut}, \eta_{cut})$, which is free
of the outgoing parton (jet).
We show corresponding $S_R(\eta_{cut})$ in Fig. \ref{fig:S_R_partonic}.
The solid line represents our partonic result. For comparison we show also
$S_R$ when only one component (valence or sea) of $F_2$ is included
in the calculation, see dashed and dotted lines. 
We see that gap survival factors for the different components are 
fairly different.
Our final result (solid line) correctly includes all components. 

The distribution of $S_R$ for the full model (solid curve) extends 
to much larger $\eta_{\rm cut}$ than the valence and sea 
contributions separately. This is due to a nonperturbative contribution, which dominates at very large
negative rapidities (see the $\eta_{\rm jet}$ distribution in 
Fig.~\ref{fig:dsigma_dyjet}). The emitted jets can be associated
only with partonic component of the model structure function. 

\begin{figure}
\includegraphics[width=.48\textwidth]{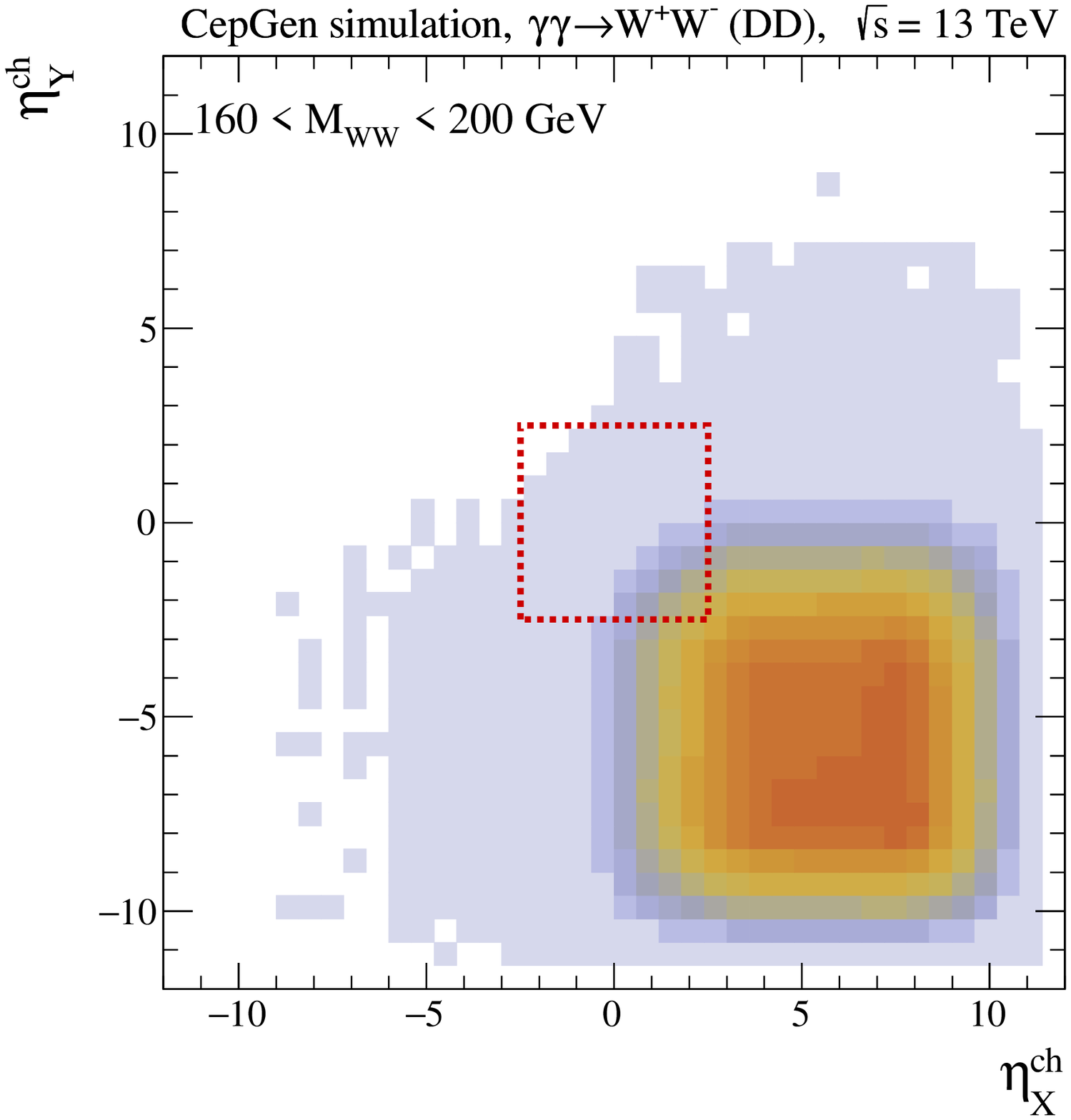}
\includegraphics[width=.48\textwidth]{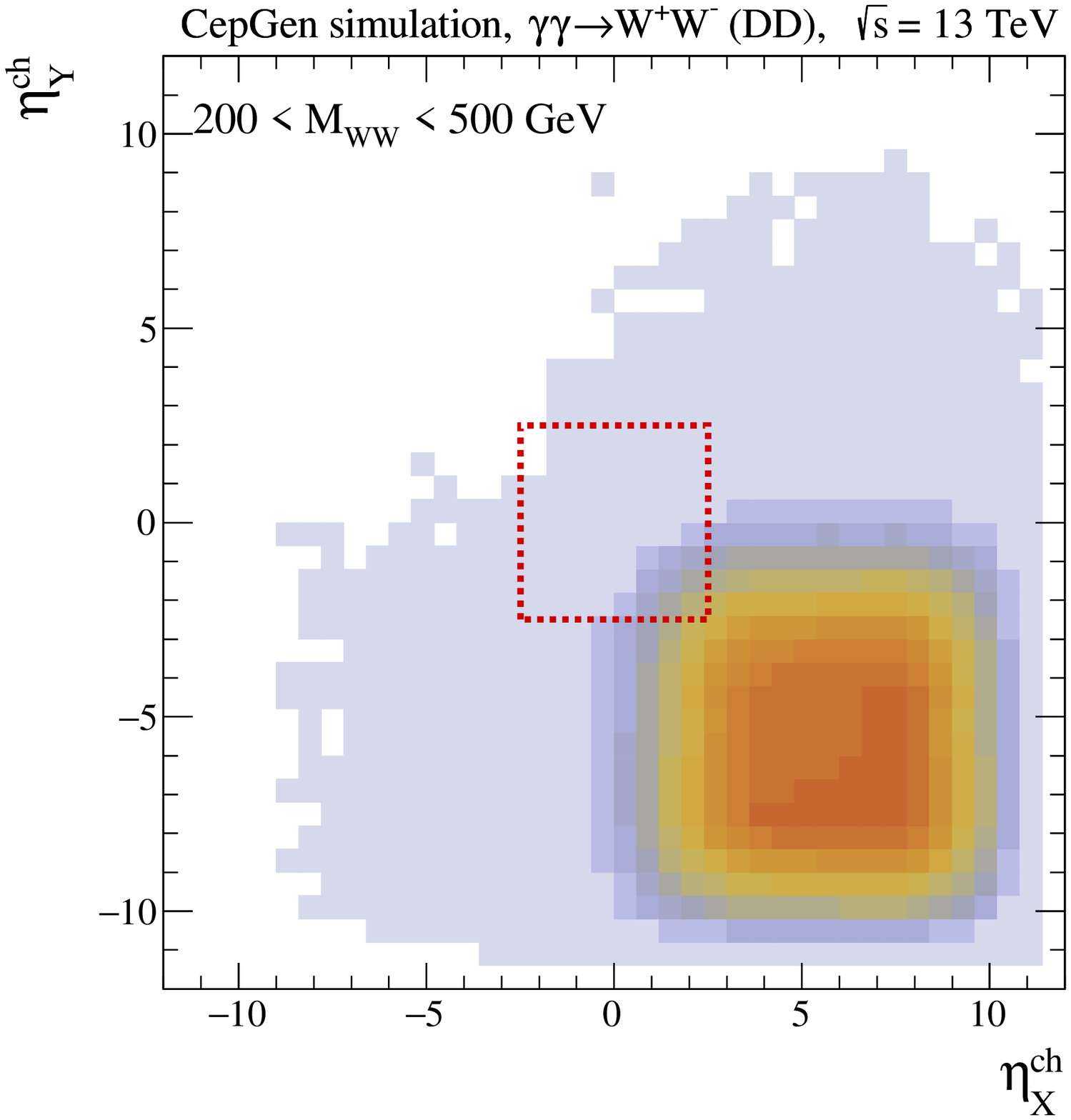}\\
\includegraphics[width=.48\textwidth]{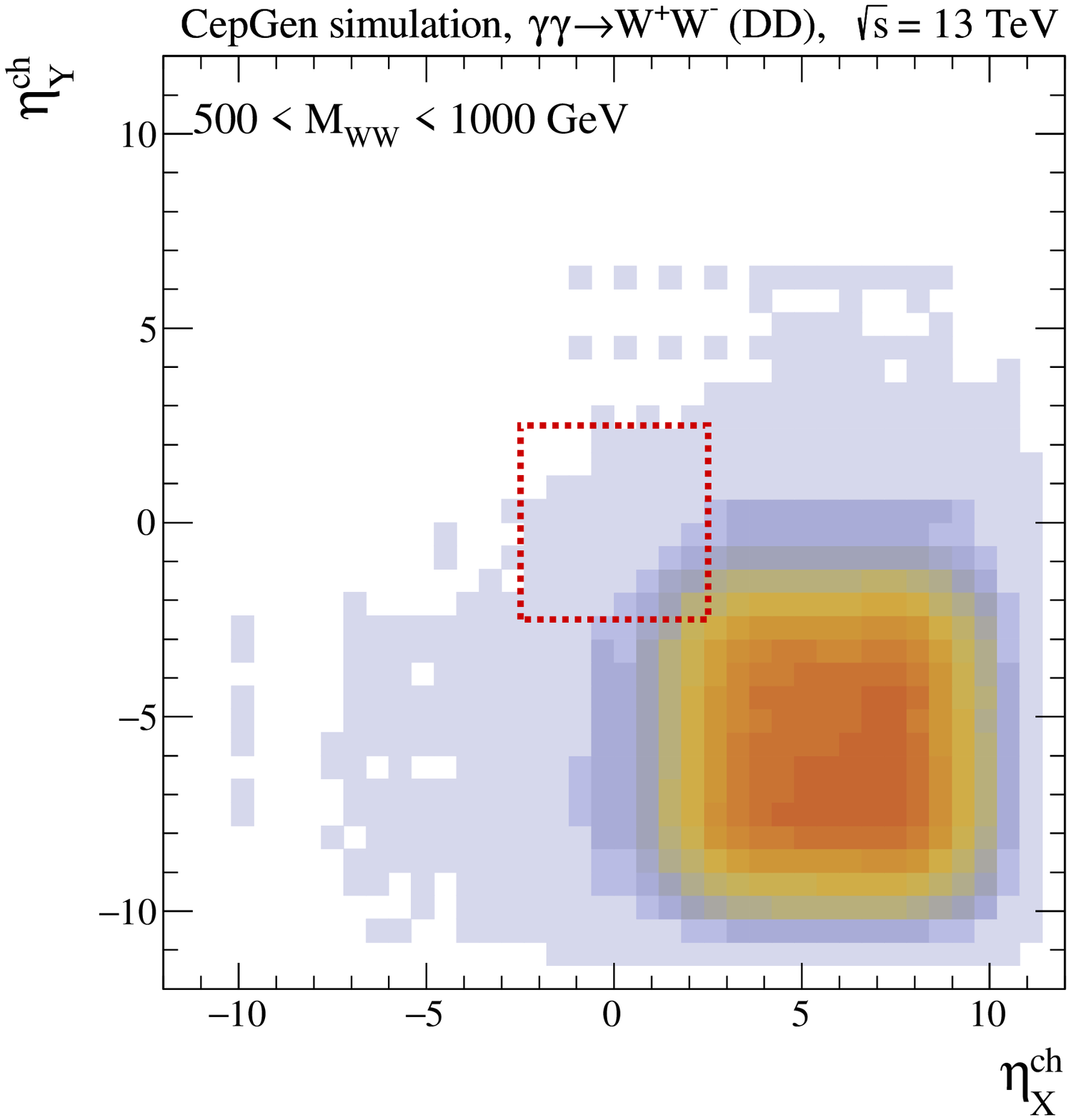}
\includegraphics[width=.48\textwidth]{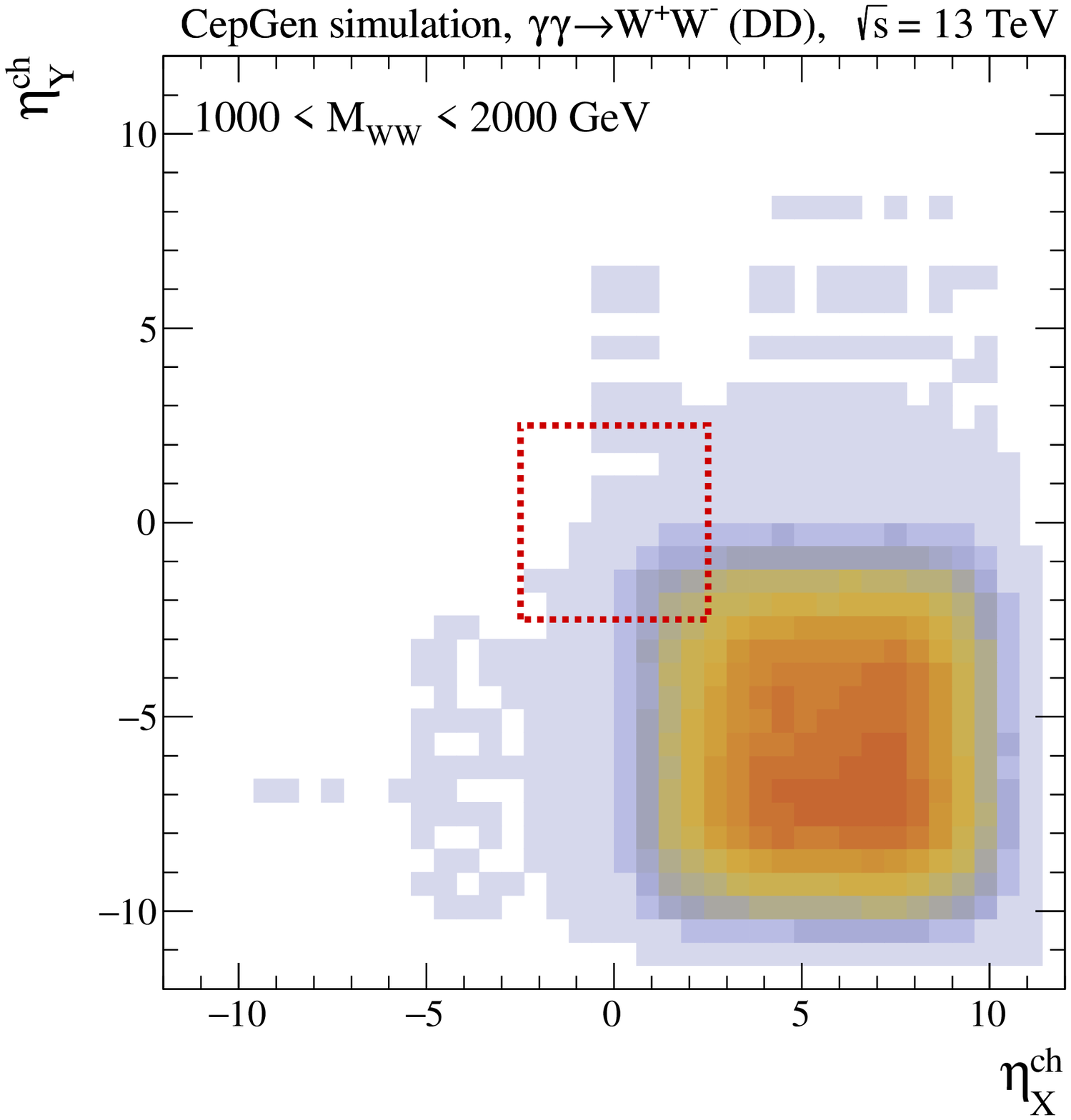}
\caption{Two-dimensional ($\eta^{\rm ch}_{X},\eta^{\rm ch}_{Y}$) distribution
for four different windows of $M_{WW}$: $(2 M_W, 200~GeV)$,
$(200, 500~GeV)$, $(500, 1000~GeV)$, $(1000, 2000~GeV)$.
The square shows pseudorapidity coverage of ATLAS or CMS inner tracker.}
\label{fig:dsig_deta1deta2_MWW_windows}
\end{figure}

In Fig. \ref{fig:dsig_deta1deta2_MWW_windows} we show two-dimensional
distributions in pseudorapidity of particles from $X$ ($\eta^{\rm ch}_X$)
and $Y$ ($\eta^{\rm ch}_Y$) for different ranges of masses of the centrally
produced system. For illustration the region relevant for ATLAS and CMS
pseudorapidity coverage is pictured by the thin dashed square.

The two dimensional plots are not sufficient to see
the dependence of the associated gap survival factor on the mass of
the centrally produced system.

We quantify this effect, see Table \ref{tab:gsf_sd}, by showing 
average remnant rapidity
gap factors for different ranges of $M_{WW}$ masses.
There we observe a rather mild dependence.
The remnant rapidity gap survival factor at fixed $\eta_{\rm cut}$
becomes larger at higher collision energies.

\begin{table}[tbp]
\centering
\begin{tabular}{|c|c|c|c|c|c|c|}
\hline
\multirow{2}{*}{Contribution} & \multicolumn{2}{c|}{$S_{R,SD}(|\eta^{\rm ch}|<2.5)$ } & \multicolumn{2}{c|}{$\left(S_{R,SD}\right)^2(|\eta^{\rm ch}|<2.5)$ } & \multicolumn{2}{c|}{$S_{R,DD}(|\eta^{\rm ch}|<2.5)$}\\
\cline{2-7}
                       & $8~\TeV$  & $13~\TeV$ & $8~\TeV$  & $13~\TeV$  & $8~\TeV$ & $13~\TeV$\\
\hline
$(2 M_{WW}, 200~\GeV)$ & 0.763(2)  & 0.769(2)  & 0.582(4)  & 0.591(4)   & 0.586(1) & 0.601(2)\\
\hline
$(200, 500~\GeV)$      & 0.787(1)  & 0.799(1)  & 0.619(2)  & 0.638(2)   & 0.629(1) & 0.649(1)\\
\hline
$(500, 1000~\GeV)$     & 0.812(2)  & 0.831(2)  & 0.659(3)  & 0.691(3)   & 0.673(2) & 0.705(2)\\
\hline
$(1000, 2000~\GeV)$    & 0.838(7)  & 0.873(5)  & 0.702(12) & 0.762(8)   & 0.697(5) & 0.763(6)\\
\hline
full range             & 0.782(1)  & 0.799(1)  & 0.611(2)  & 0.638(2)   & 0.617(1) & 0.646(1)\\
\hline
\end{tabular}
\caption{Average rapidity gap survival factor related to remnant fragmentation
	for {\it single dissociative} and {\it double dissociative} contributions
for different ranges of $M_{WW}$.
All uncertainties are statistical only.
}
\label{tab:gsf_sd}
\end{table}

\begin{figure}
  \centering
  \includegraphics[width=.48\textwidth]{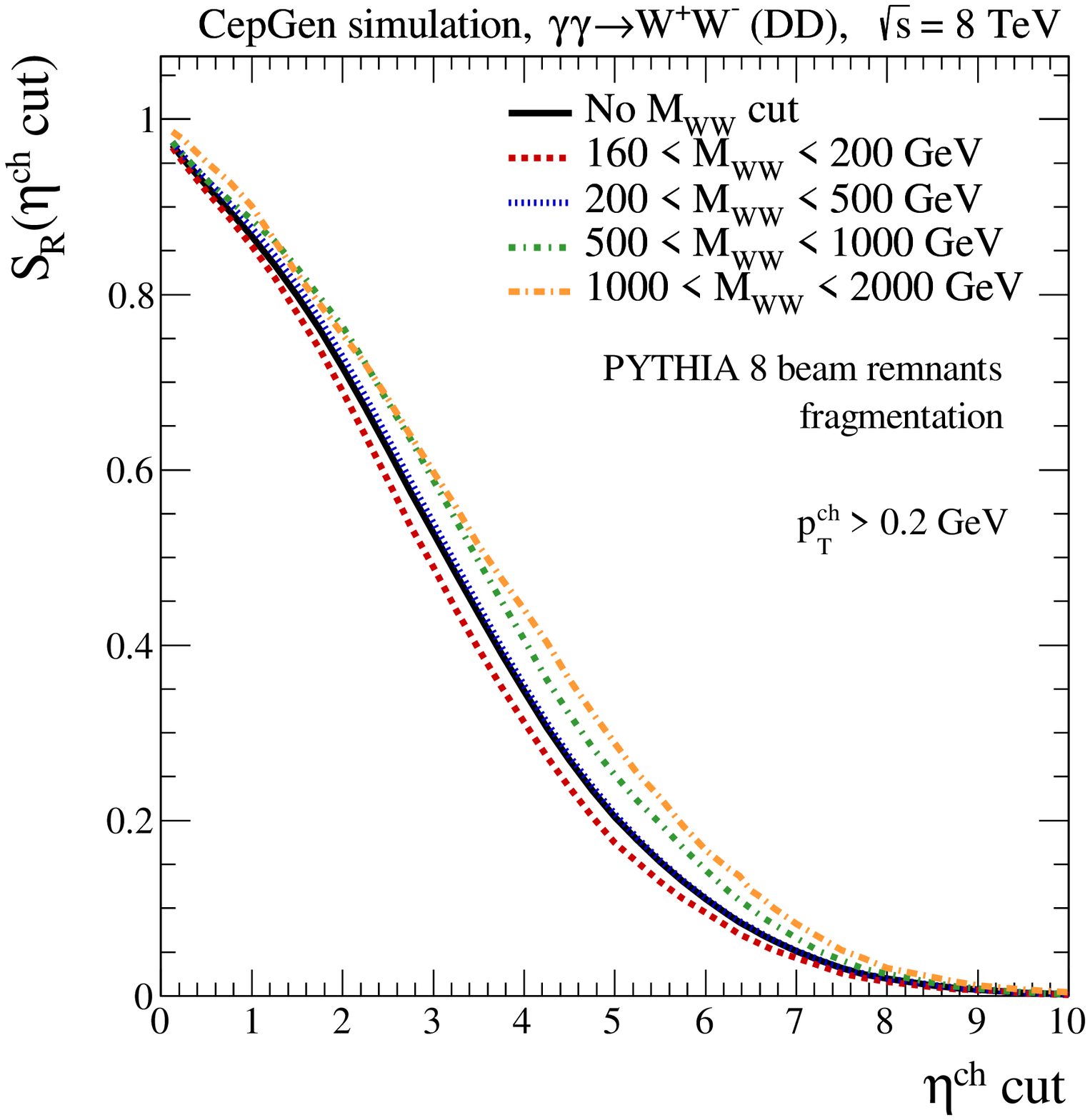}
  \includegraphics[width=.48\textwidth]{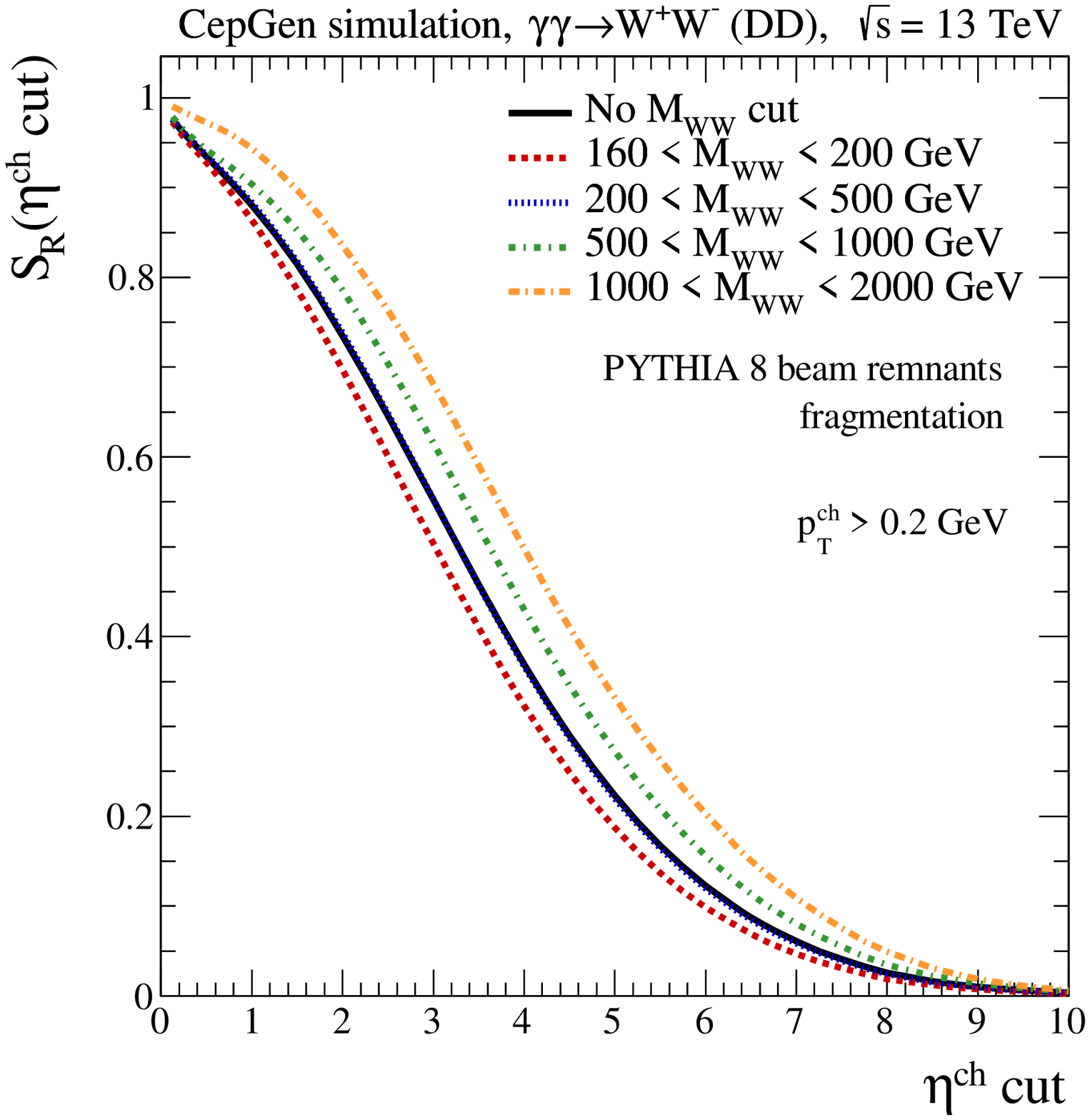}
  \caption{Gap survival factor for double dissociation as a function of
    the size of the pseudorapidity veto applied on charged particles emitted from
    proton remnants, for the
    diboson mass bins defined in the text and in the figures for $\sqrt{s} = 8~\TeV$ (left)
    and $13~\TeV$ (right).}
  \label{fig:surv_vs_etacut_dd}
\end{figure}
In Fig. \ref{fig:surv_vs_etacut_dd} we show the distribution in $\eta_{\rm cut}$ for the double
dissociation process. We predict a strong dependence on $\eta_{\rm cut}$. It would be
valuable to perform experimental measurements with different $\eta_{\rm cut}$.

\section{Results for $t \bar t$ pairs production}

In Table \ref{tab:sig_tot} we show integrated cross sections for each
of the categories of $\gamma \gamma$ processes shown in Fig.\ref{fig:diagrams}.
We observe the following hierarchy as far as the integrated
cross section is considered:
\begin{equation}
\sigma_{t \bar t}^{\rm el-el} < \sigma_{t \bar t}^{\rm in-el} 
= \sigma_{t \bar t}^{\rm el-in} < \sigma_{t \bar t}^{\rm in-in}.
\label{eq:sigma_hierarchy}
\end{equation}
The summed inclusive cross section at $\sqrt{s} = 13 \, \rm{TeV}$ 
is 2.36 fb. This is a rather small number in comparison with other inclusive production mechanisms. In the right panel of Table \ref{tab:sig_tot} we show results when 
a rapidity gap 
\footnote{that means no additional particle production except $t$ or $\bar t$} 
in the central region, for $-2.5 < y <2.5$ is required in addition.
In principle, imposing this condition 
requires modelling of the full final state, as we did 
for the case of $W^+ W^-$ production.
As in each event we have the full four-momentum of the virtual photon(s),
as well as the invariant masses of the proton remnants, the four-momenta
of the recoiling jet(s) can be reconstructed.
To a good accuracy the rapidity gap condition is equivalent
to require that the recoiling jets fulfill $|y_{\rm jet}| > 2.5 $.

\begin{table}[tbp]
\centering
\begin{tabular}{|c|c|c|}
\hline
Contribution          &  No cuts & $y_{\rm jet}$ cut\\
\hline
elastic-elastic       &  0.292 &  0.292 \\
elastic-inelastic     &  \multirow{2}{*}{0.544} &  \multirow{2}{*}{0.439} \\
inelastic-elastic     &  & \\
inelastic-inelastic   &  0.983 &  0.622 \\
\hline
all contributions     &  2.36  &  1.79 \\
\hline
\end{tabular}
\caption{Cross section in fb at $\sqrt{s}$ = 13 TeV for 
different components (left column) and the same when the extra condition
on the outgoing jet 
$ |y_{\rm jet}| > 2.5$
is imposed.}
\label{tab:sig_tot}
\end{table}

\begin{figure}
  \centering
  \includegraphics[width=.48\textwidth]{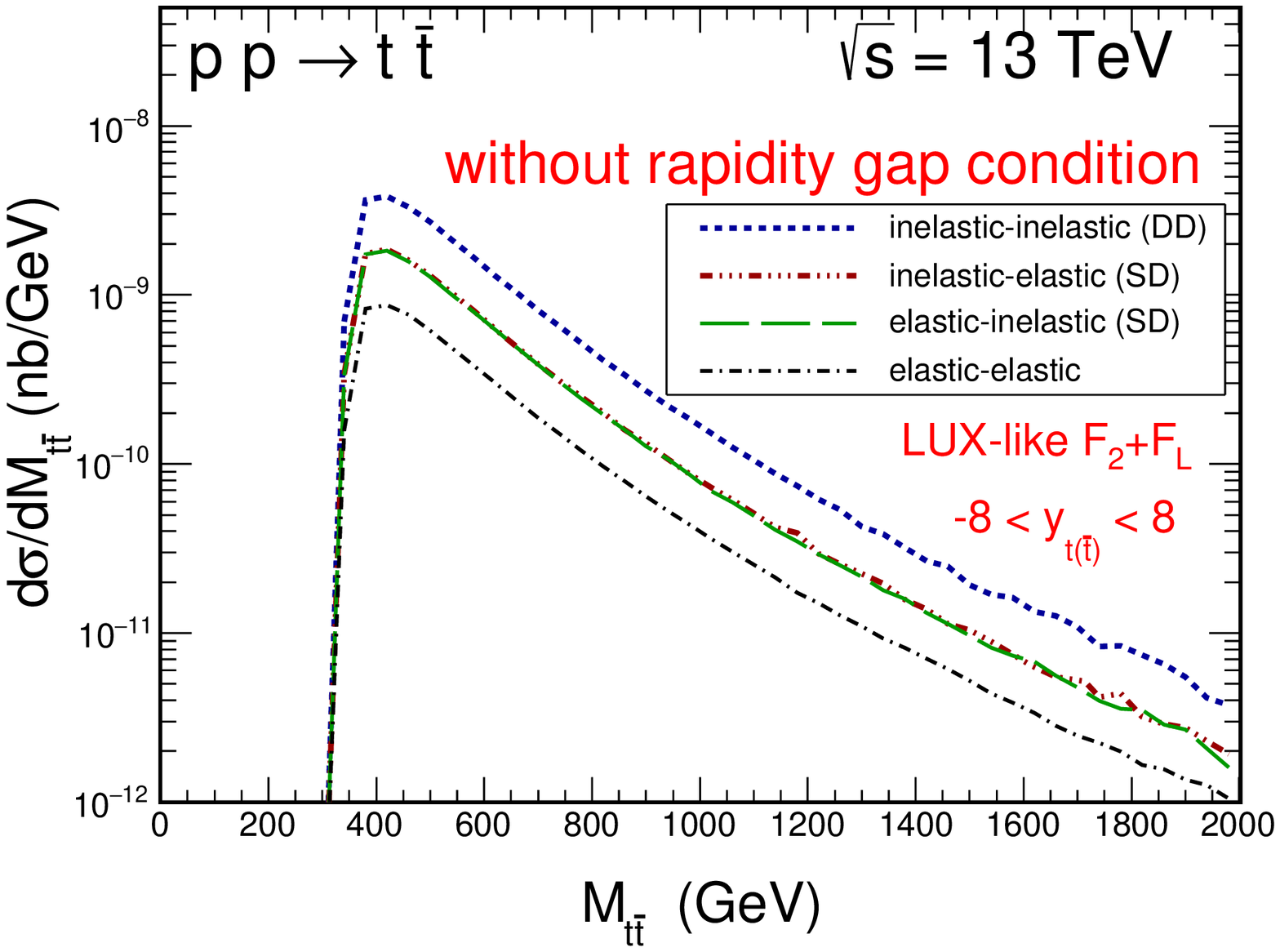}
  \includegraphics[width=.48\textwidth]{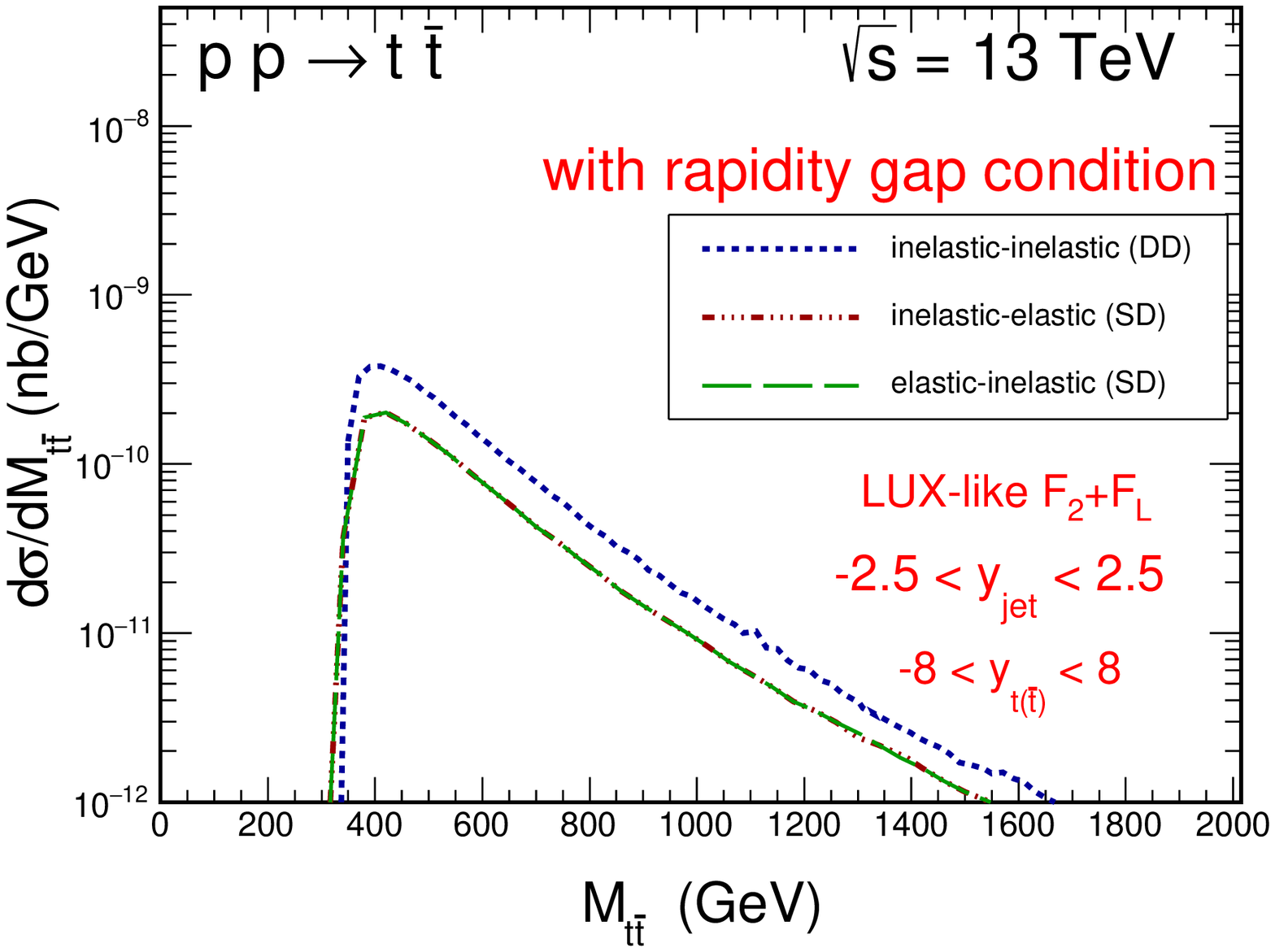}
  \caption{$t \bar t$ invariant mass distribution for different 
components defined in the figure. The left panel is without imposing
the condition on the struck quark/antiquark and the right panel includes
the condition.
}
\label{fig:dsig_dMttbar}
\end{figure}

The same is true for the distribution in $t \bar t$ invariant mass
(see the left panel of Fig.\ref{fig:dsig_dMttbar}). The distributions 
are almost identical and differ only by normalisation.
In the right panel of Fig.\ref{fig:dsig_dMttbar} we show
similar results when conditions on outgoing light 
quark/antiquark jets are imposed. 
The extra condition leads to a
lowering of the cross section with only very small modification of 
the shape of the $M_{t \bar t}$ distribution.

\begin{figure}
  \centering
  \includegraphics[width=.48\textwidth]{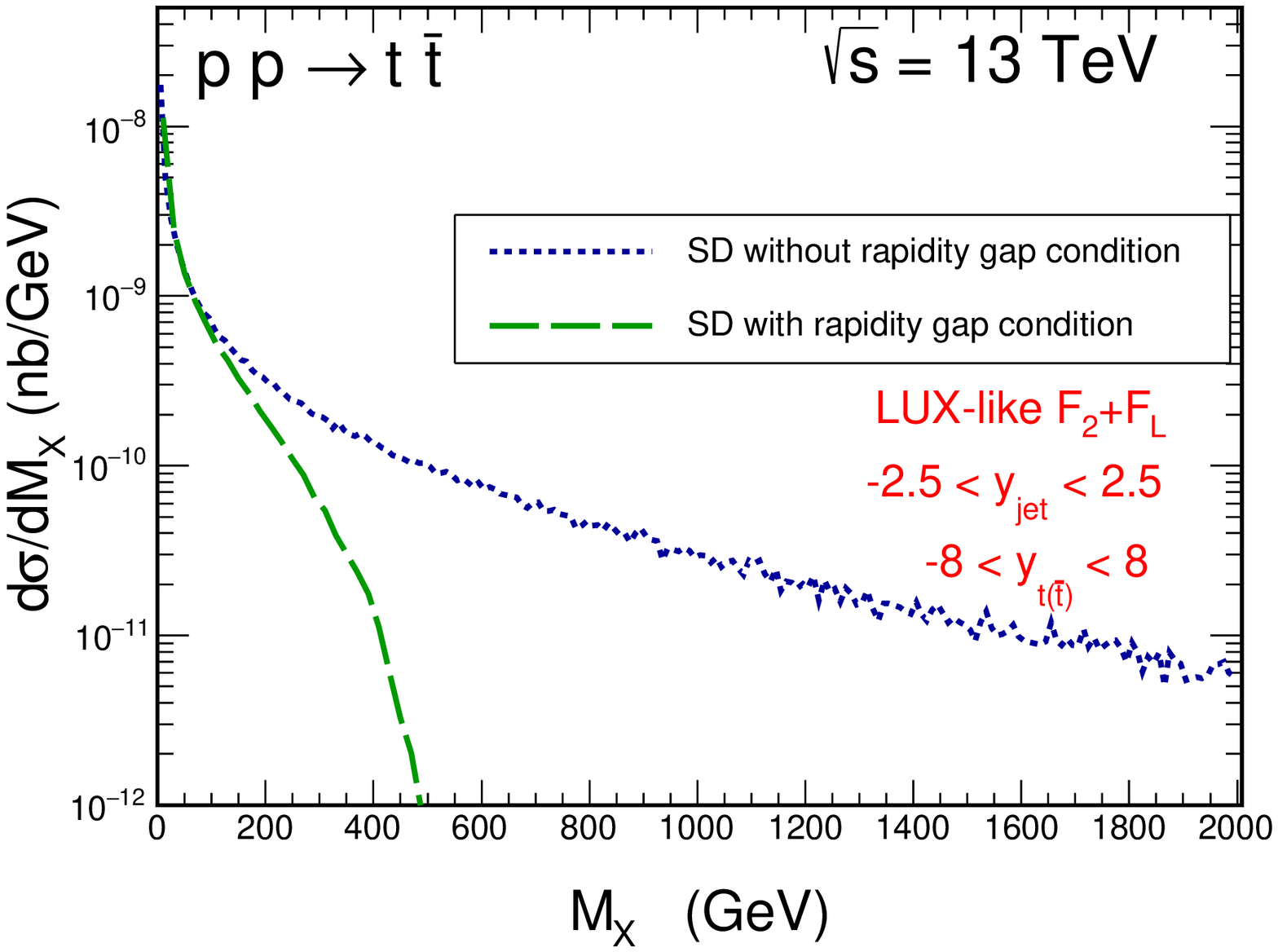}
  \includegraphics[width=.48\textwidth]{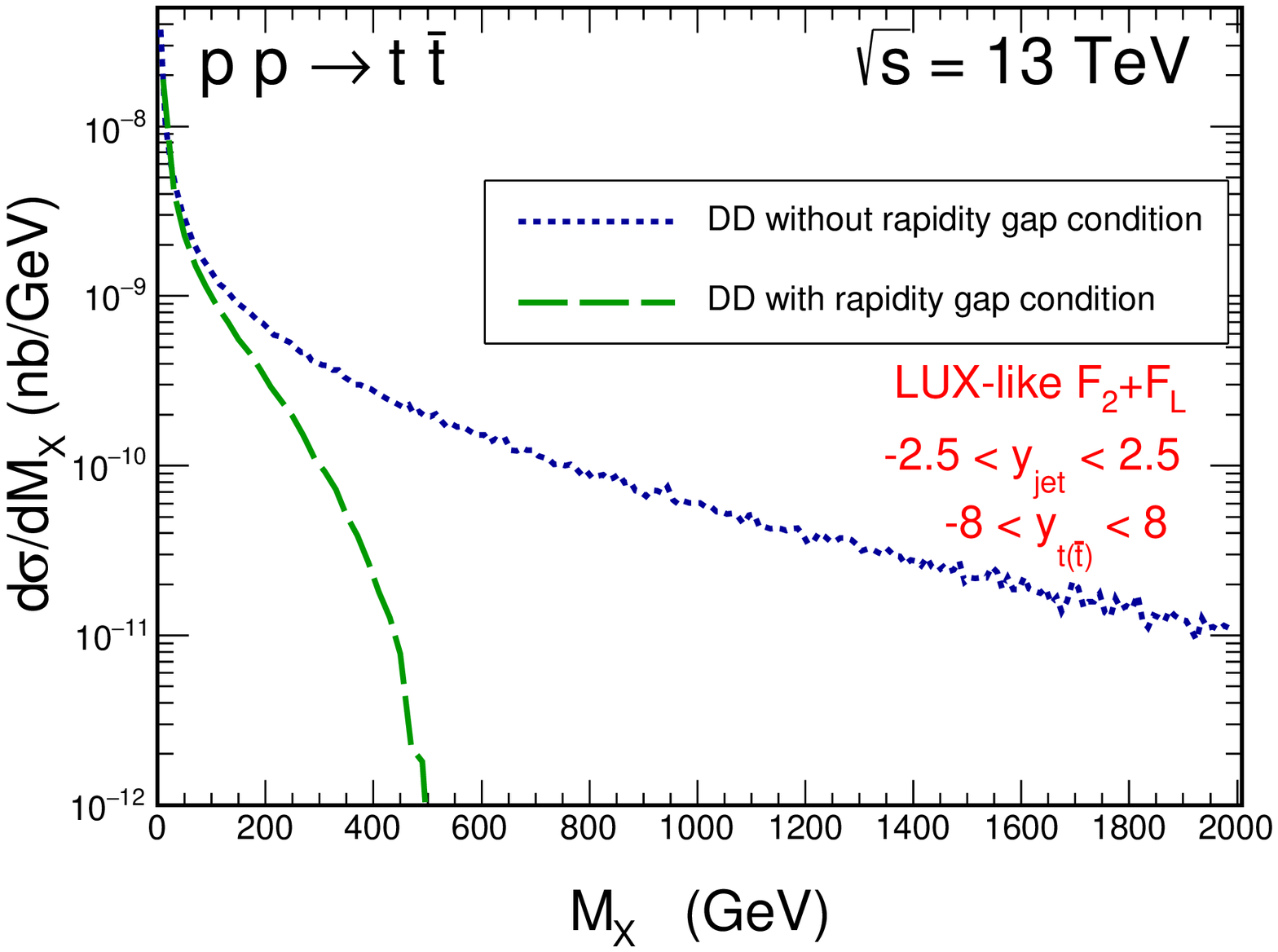}
    \caption{Distribution in the mass of the dissociated system 
for single dissociation (left) and double dissociation (right).
We show result without and with the rapidity gap condition.
}
\label{fig:dsig_dMX}
\end{figure}

In addition, in Fig.\ref{fig:dsig_dMX} we show distributions 
in outgoing proton remnant masses $M_X$ and/or $M_Y$. Similar shapes are observed for
single-dissociative and double-dissociative processes.
Population of large $M_X$ or $M_Y$ masses is associated with the
emissions of jets visible in central detectors 
(i.e. with -2.5 $< y_{\rm jet} <$ 2.5).
We show the distribution in the remnant mass $M_X$ separately for the single dissociation (left) and double
dissociation (right).
As can be seen, the rapidity gap requirement introduces a rather sharp 
cut-off in the large-mass tail of the $M_X$-distribution.

\begin{figure}
  \centering
  \includegraphics[width=.48\textwidth]{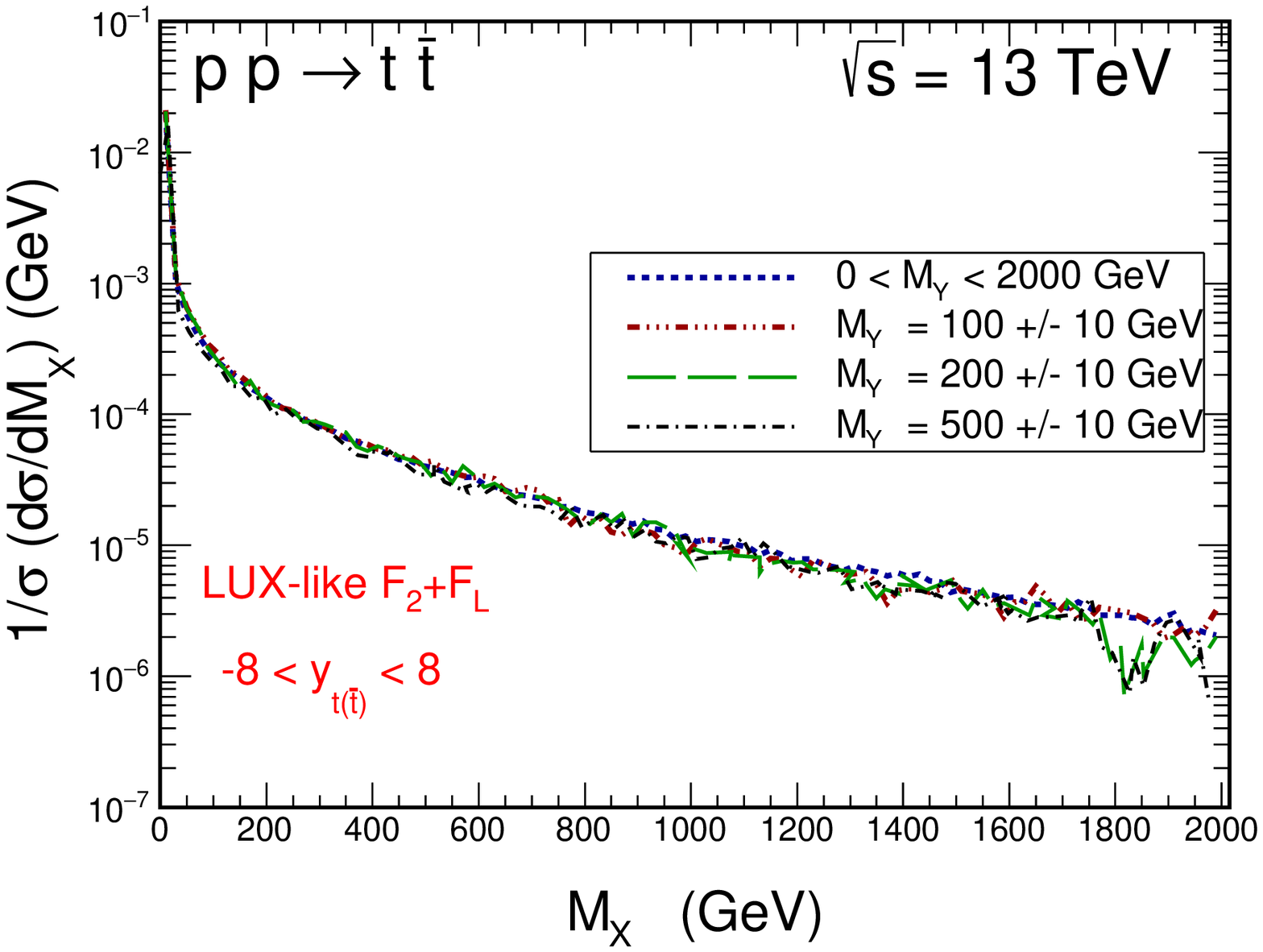}
  \includegraphics[width=.48\textwidth]{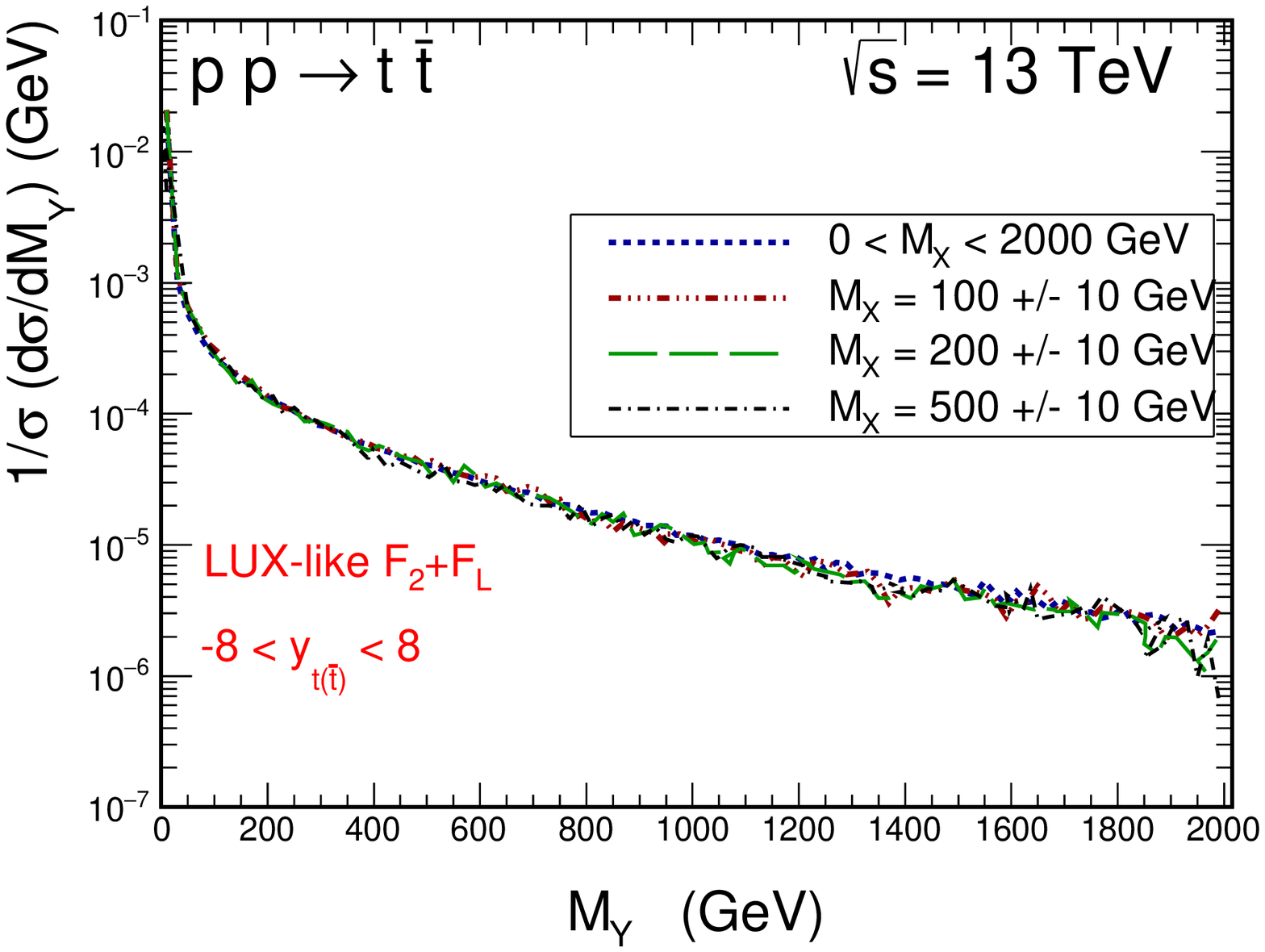}
    \caption{Distribution in $M_X$ for different windows of $M_Y$ (left)
and as a function of $M_Y$ for different windows of $M_X$ (right).
}
\label{fig:factorization}
\end{figure}

In Fig.\ref{fig:factorization} we show distributions in
$M_X$ for a fixed $M_Y$ (left panel) and in $M_Y$ for a fixed $M_X$
(right panel). The distributions are arbitrarily normalized
to the same integral. All the distributions coincide.

\begin{figure}
  \centering
  \includegraphics[width=.50\textwidth]{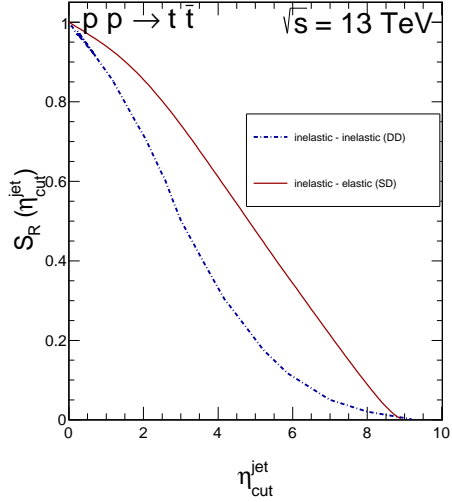}
  \caption{Gap survival factor for single and double dissociation as a function of
    the size of the pseudorapidity veto applied on the recoiling jet emitted from
    proton remnants.
}
\label{fig:s_r_dd}
\end{figure}

Finally, in Fig.\ref{fig:s_r_dd} we show our results for
$pp \to \gamma \gamma \to t \bar t$ processes. The gap survival
factor fullfills the relation $S_R^{DD} < S_R^{SD}$. We have checked 
that the factorisation $S_R^{DD} = (S_R^{SD})^2$ holds to very good accuracy.

\section{Conclusions}
In this presentation we have discussed the quantity called
remnant gap survival factor for the $pp \to W^+ W^-$ and $pp \to t \bar t$ reactions
initiated via photon-photon fusion.
We use our formalism developed
for the inclusive case \cite{Luszczak:2018ntp} which includes
transverse momenta of incoming photons. The partonic formalism has been supplemented
by including remnant fragmentation that can spoil the rapidity gap
usually used to select the subprocess of interest.
We quantify this effect by defining the remnant gap survival factor
which in general depends on the reaction, kinematic variables and
details of the experimental set-ups.
We have found that the hadronisation only mildly modifies
the gap survival factor calculated on the parton level. We find different values for double and single
dissociative processes. In general, $S_{R,DD} < S_{R,SD}$
and $S_{R,DD} \approx (S_{R,SD})^2$.

The cross sections for production of $t \bar t$ pairs via $\gamma^* \gamma^*$
fusion summed over the different categories of processes 
is about 2.36 fb (full phase space), i.e. rather small compared to
the standard inclusive $t \bar t$ cross section (of the order of nb).
Our results imply that for the production of such heavy objects as $t$ quark
and $\bar t$ antiquark the virtuality of the photons attached to 
the dissociative system are very large ($Q^2 <$ 10$^{4}$ GeV$^2$). 


\end{document}